# A random phased-array for MR-guided transcranial ultrasound neuromodulation in non-human primates

## Authors:

**1. Vandiver Chaplin   (*)**
Vanderbilt University Institute of Imaging Science
1161 21st Avenue South
Nashville, TN 37232
**\*corresponding author**

2. **Marshal A. Phipps**
Vanderbilt University Institute of Imaging Science

**3. Charles F. Caskey**
Vanderbilt University Institute of Imaging Science

Department of Radiology and Radiological Sciences
Vanderbilt University Medical Center
1161 21st Avenue South
Nashville, TN 37232

## Abstract

Transcranial focused ultrasound (FUS) is a non-invasive technique for therapy and study of brain neural activation. Here we report on the design and characterization of a new MR-guided FUS transducer for neuromodulation in non-human primates at 650kHz. The array is randomized with 128 elements, radius of curvature 7.2cm, and opening diameter 10.3cm (focal ratio 0.7). Simulations were used to optimize transducer geometry with respect to focus size, grating lobes, and directivity. Focus size and grating lobes during electronic steering were quantified using hydrophone measurements in water and a three-axis stage. A novel combination of optical tracking and acoustic mapping enabled measurement of the 3D pressure distribution in the cortical region of an ex vivo skull to within ~3.5 mm of the surface, and allowed accurate modelling of the experiment via non-homogeneous 3D acoustic simulations. The data demonstrates acoustic focusing beyond the skull bone, with the focus slightly broadened and shifted proximal to the skull. The fabricated design is capable of targeting regions within the S1 sensorimotor cortex of macaques.

# Introduction

Focused ultrasound (FUS) is a promising method for non-invasive neural stimulation in the brain. Early studies have shown that neurological conditions such as essential tremors can be treated with ultrasound ablation (Elias et al. 2016), while sub-thermal stimulation is being explored to probe functional connections in the brain and may have therapeutic benefits in a variety of neurological diseases (Panczykowski et al. 2014; Lee, Kim, et al. 2016). Functional experiments typically use a single-element transducer, placed over a target with a mechanical positioning system and use MRI, optical tracking, or visual estimation for transducer alignment (Lee, Chung, et al. 2016; Deffieux et al. 2013; King et al. 2014). An array-based system offers advantages such as electronic steering, which is typically feasible within a 3-4 cm of the natural focus, depending on transducer geometry (Malietzis et al. 2013; Hand et al. 2009; Hynynen et al. 2004; Payne et al. 2011). An array also enables correction of phase-aberrations due to the skull (Larrat et al. 2010; Vyas et al. 2012; Marsac et al. 2012; Kaye et al. 2012). Although array-based systems for transcranial ablation exist for human subjects, they are not optimized for cortical stimulation, which requires focusing closer to the skull than the deep-brain targets reported previously. Nor have arrays been designed to target the smaller neural anatomy of non-human primates. Here we report on the design and characterization of a new spherically-focused randomized sparse array transducer for selectively targeting the sensorimotor cortex (S1 region) of macaques.

Much of the work on ultrasound 2D array design has come from the volumetric imaging community, where large-aperture 2D arrays are needed for quality image resolution at depth. To achieve 1˚ or better angular resolution, a 60λ aperture is needed, but to avoid large grating lobes element spacing must be less than λ/2, necessitating a prohibitively large number (~$10^4$) of elements to fill the aperture (Martínez et al. 2010). Subsequent research into sparse 2D array configurations that minimize grating lobes has led to a variety of fruitful results, such as Vernier arrays that use two different sets of elements for transmit and receive that are spatially periodic and offset, Fermat spirals, and randomized arrays (Lockwood et al. 1996; Martínez et al. 2010). These studies focus on factors affecting image quality such as side lobe and grating lobe levels, and azimuthal beam symmetry. Generally spirals have lower grating lobe artifacts in image reconstruction, followed by Vernier and randomized, though the differences are small and the results seem to depend on aperture shape and the apodization function used. The ideal design for reducing grating lobes is a dense array with near 100% coverage over the aperture and center-to-center element spacing below λ/2, but the cost is prohibitive with current fabrication techniques.

In the case of focused therapeutic arrays, the acoustic design problem is simpler since it is restricted to the one-way (transmit) radiation pattern, rather than the convolution of transmit and receive patterns encountered in acoustic imaging. However, larger transmit power and integration with image-guidance pose unique hardware challenges. Image-guidance is important for accurate spatial targeting and typically requires MR-compatible hardware, integration with co-registered ultrasound probes, registration-based tracking in CT or MRI volumes, or some combination of these methods (Ebbini et al. 2006; Wang et al. 1994; Vanbaren et al. 1996; Kim et al. 2012). Secondary maxima in the transmit field (often synonymous with grating lobes in focused ultrasound literature) can cause heating in non-targeted regions in hyperthermia. FUS array design has therefore been concerned with maximizing power deposition and steering capabilities while minimizing off-target heating, using acoustic simulations coupled with thermal models (Stephens et al. 2011; Clement et al. 2000; Ellens et al. 2011; Daum & Hynynen 1999; Hughes et al. 2016; Ebbini & Cain 1991). Focusing of ultrasound with a small number of elements for therapeutic purposes dates to the mid-20$^{th}$ century with the experiments of Lynn, Putnam, Fry, and others (Lynn & Putnam 1944; Fry et al. 1954). Of particular importance is the advent of placing transducer elements on a spherical cap (a.k.a., spherical shell, spherical section)

(Ebbini & Cain 1991). Irrespective of element configuration, constraining elements to a spherical surface reduces the amplitudes of secondary maxima relative to the focus but limits off-axis focusing.

Spherically-focused arrays have seen widespread development among FUS research groups and commercial ventures. Requirements such as large acoustic power output and lower frequencies generally necessitate larger element sizes than encountered in imaging. The simultaneous constraints of a large radius of curvature for non-invasive deep tissue sonication, and a focal ratio between 0.5 – 1.0 for effective focal gain without grating lobes lead to large-area transducers. In 1999 Daum and Hynynen presented a 256-element spherical cap array for ablation, with a 10cm radius of curvature and focal ratio 0.83. The surface area was near 100%-covered with equal-area elements, element spacing was nearly periodic and selected to allow safe electronic steering within a 1x1 cm^2 area around the natural focus while maintaining grating lobes < 10% of the focal pressure (Daum & Hynynen 1999). In 2004 a similar but larger, 500-element dense spherical array was presented for the purpose of human transcranial ablation of deep-brain targets (Hynynen et al. 2004). This system had a similar focal ratio but larger radius of curvature (15cm) to reduce incident energy density across the skull bone, and a large number of elements to maximize correction of phase aberrations, inferred from trans-cranial simulations (Clement et al. 2000).

Sparse spherical arrays have also received significant attention as a way to achieve large-aperture capabilities with a cost-effective number of elements. In 1996 Goss et *al.* presented an early sparse array for focused ultrasound surgery, examining how array packing may effect performance as the beam is steered off-axis (Goss et al. 1996; Frizzell et al. 1996). Since then studies have been conducted for a range of therapeutic array designs considering both element arrangement and size. Gavrilov and Hand used simulations to demonstrate that randomized arrays have lower grating lobes than regularly ordered arrays, and that larger elements decrease array steering performance (Gavrilov & Hand 2000). A similar result was shown by Ellens *et al.* in a simulation study coupled to Pennes' bio-heat equation, where they demonstrated that grating lobe heating increases for arrays with more sparsity (i.e., less coverage) (Ellens et al. 2011). Stephens *et al*. presented a comprehensive simulation study comparing spherical and annular arrays having randomized, periodic, and spiral element arrangements (Stephens et al. 2011). Among the conclusions are that spherical arrays reduce grating lobes when elements are randomized, instead of using a periodic arrangement. Furthermore, if the same active area is divided among a small number of large elements, vs. a large number of small elements, the transducer has a natural focus that is smaller and closer to the geometric radius, at the cost of off-axis steering performance. Such an array is said to have greater directivity.

Sparse annular arrays have also been designed, typically to hold a second device in the central void space, such as a focused transducer that enables multi-frequency sonications, or an imaging probe for co-registered FUS and ultrasound imaging (Ji et al. 2011; Ebbini et al. 2006; Hand et al. 2009). These arrays have similar properties to full spherical caps but can have slightly elevated axial and side lobes, increasing risk for off-target heating in thermal therapy (Stephens et al. 2011). This problem has been recently addressed by using a spiral element configuration rather than a randomized one. Ramaekers et *al.* explored various Fermat-spiral designs and tested one using the golden angle relation (the "sunflower pattern") on a spherical cap (Ramaekers et al. 2017). While there was no direct comparison to a randomized array presented, the performance was qualitatively good and may offer advantages in a tighter focus and larger output than a random array owing to its high area coverage. Overall, spherical sparse arrays are good candidates for therapeutic transducer design, but array size, sparsity, element count, and element arrangement offer tradeoffs between cost, grating lobes, focus size, and steering performance that should be considered per application.

The general problem of packing equally-sized circular elements on a sphere without overlap, as required for our randomized array design, is well-studied. The key result is that maximal packing density is a function of the number of circles (arbitrarily sized to fit on a unit sphere), with 90.6% being the theoretical limit (an infinite number of elements). Exact packing solutions have been discovered for N=2, 12, and 24 elements, but in practice numerically optimized solutions are used for arbitrary numbers that result in typical densities ~85% (Clare & Kepert 1991). However, these are highly ordered configurations and thus if used as the basis for an array design would be subject to the λ/2 spacing constraint to avoid grating lobes. If array density above 90% is desired, non-circular elements must be used, of which there are several examples (Ramaekers et al. 2017; Hynynen et al. 2004).

Given the success of recent spherical random arrays across in both transcranial and abdominal applications, and the established technique for their fabrication, we chose to pursue a sparse cap design to achieve our goal of selectively targeting the macaque S1 region. Our design problem was simplified by the choice of 650kHz as the array center frequency, based on the frequency optimization work of Clement, the success of 650kHz in transcranial human studies, and that 650kHz has been shown to modulate neurons (Clement et al. 2000; Kim et al. 2014). We fixed the element diameter to 6.6 mm to maintain a low electrical impedance and high acoustic power output at 650 kHz. With element size and center frequency fixed, our design goal was to optimize a specific spherical geometry and channel count capable of focusing and steering over the macaque S1 region transcranially with sidelobes that would not generate off-target stimulation. We developed a GPU-based Rayleigh-Sommerfeld implementation to simulate the free field while varying element number, coverage, and geometry, arriving at a final design with radius of curvature (ROC) of 72mm, diameter 103 mm, and 128 channels. The chosen design is an approximate midpoint among various tradeoffs in array directivity, size, acoustic power output, and cost. Below we present both the simulation studies and characterization of the fabricated transducer output with steering. Transcranial beam maps behind a sample *ex vivo* macaque skull are also presented.

## Methods

The overall design objective was to develop an array capable of steering over the macaque somatosensory cortex (S1), with a focus small enough to modulate S1 in general, and possibly subregions associated with various neural circuits, such as face or hand stimulation. In adult macaques the S1 anatomical region is typically ~2cm long, has a roughly 3mm cross-section and is ~1cm from the interior of the skull bone. Though the precise spatial relationship between the extent of an acoustic focus and regions of neural modulation is an open question, we used the S1 anatomical size to inform the desired focus size of our design, leading to design requirements that focus width should be less than 3mm and length about 1cm, to effectively target portions of S1 (Robertson et al. 2017). Beam shape, steering and grating lobes were evaluated with free-field Rayleigh-Sommerfeld simulations.

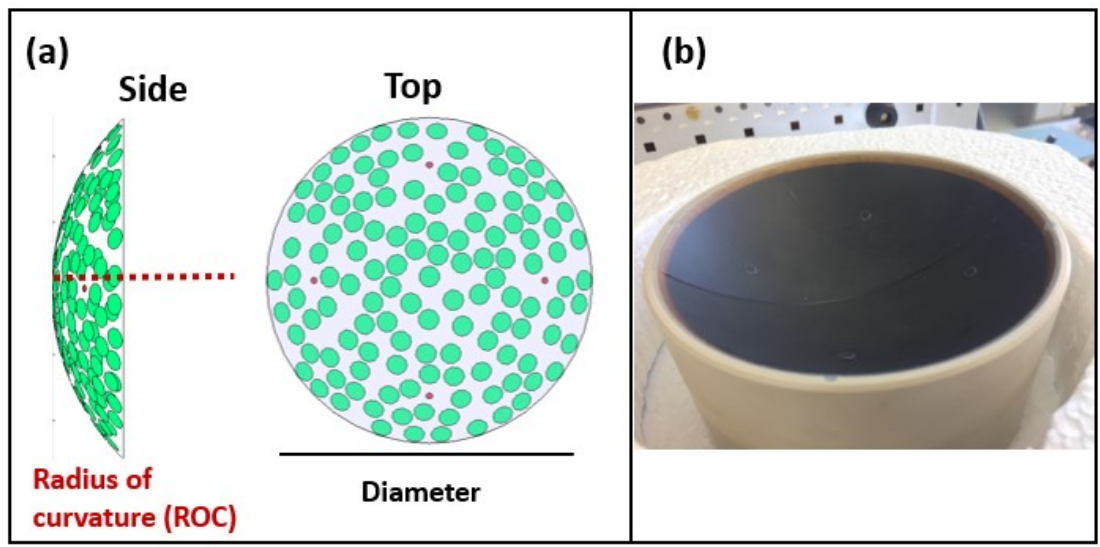

**Figure 1 – (a)** Setup of randomized spherical caps tested with simulations. ROC and diameter were varied with fixed element sizes. The green circles are the positions of the 128 transmit elements in their final arrangement. Red circles show the location of passive cavitation detectors. **(b)** Photo of the fabricated design.

*Design Simulations*

To optimize transducer geometry with respect to focus size and steering, we simulated a subset of the design variable space (radius of curvature, diameter, and coverage). The transducer shape used was a solid spherical cap with circular elements placed on the surface. Elements were randomly placed in each iteration, and were simulated using a superposition of 60 point sources per element conforming to the convexity of the spherical surface. Element sizes were fixed, and the number of elements was the largest number of 64, 96, 128, 160 that could fit onto the transducer in a given geometry without overlap and a minimum margin of 1 mm (i.e., an effective diameter of 7.6mm). In addition to the transmit elements, four passive cavitation detectors (PCDs) 2.2mm in diameter were inserted at fixed points, and random element placement was also constrained to avoid these. **Figure 1** shows the geometric setup used for simulations.

The effect of area coverage was evaluated independently to verify the general proposition that grating lobes decrease with increasing coverage. To study coverage, a specific geometry was chosen, the number of elements set to 128, and the element size was varied such that effective coverage ranged from 10% to 100%. The number of point sources approximating each element varied so that the density of point sources was roughly held constant, and element boundaries were allowed to overlap. This method, while not feasible to fabricate due to element overlap, allows evaluating coverage ratios above the packing limit, while also maintaining pseudo-randomization in point source placement. Without element overlap, for 128 randomly placed elements of 6.6 mm diameter 1 mm margin, the maximum coverage obtained was 49% in the geometries tested.

We implemented the Rayleigh-Sommerfeld integral in a GPU kernel using Nvidia's CUDA C++ API for all free field simulations. The homogenous-medium integral is an ideal use-case for massive parallelization, since there is no need for communication between threads, minimal conditional logic required, and each output voxel in the simulation grid can be computed independently of other voxels. One GPU thread was allocated for each voxel in the simulation grid. Within each thread, the contribution from all *N* point sources comprising the sub-sampled array elements (each point having an amplitude and phase represented as a complex number) was computed. Element positions, normal vectors, complex array amplitudes, and voxel positions were loaded into GPU constant memory, and one thread per output voxel was executed. The kernels were wrapped in a Python interface, and simulations were setup and run in Python, passing necessary parameters via NumPy arrays. Code and instructions to build the GPU library are available on Github (doi.org/10.5281/zenodo.1112442). The repository also contains functions to generate randomized arrays.

*Fabrication*

The chosen design was fabricated by Imasonic (Besancon, France) in two spherical sections with elements diced out of the piezocomposite material. Image Guided Therapy (Pessac, France) built the electronic generators that power the transducer, and 650kHz electrical matching box. The entire system is MR compatible.

*Output*

A three-axis motion stage was used to translate a hydrophone for mapping acoustic output. Output at low power was measured using a needle hydrophone (HNC-400, Onda Corporation), and an optical hydrophone (Precision Acoustics Ltd, Dorchester, UK) at high power. With the needle hydrophone, beam maps and steering profiles were collected. The beam was electronically steered in 10-mm steps in both the transverse and axial directions in order to quantify grating lobe levels. Peak pressures at the focus were collected with the optical hydrophone as power was increased to 100% output. Hydrophone signals were digitized and recorded with a programmable oscilloscope controlled via Python.

*Transcranial FUS*

To evaluate a transcranial sonication, beam maps in a cortical region behind an *ex vivo* skull piece were collected using a novel approach. Prior to measurement, the skull was placed in degassed water under a vacuum for 24 hours. Because of skull attenuation and the generally lower pressures used for neuromodulation, measuring acoustic fields behind the skull is best with the sensitivity of a needle hydrophone. However the needle tip is susceptible to damage, making it a challenge to safely navigate the hydrophone in close proximity to the skull (ie. the cortical region we wish to stimulate).

To solve this problem, we used a combination of optical tracking and image registration. First, a CT of the skull piece was collected (vivaCT 80 Scanco Medical, Switzerland). The skull piece was then rigidly mounted to a water tank several centimeters from the transducer face. An NDI Polaris optical tracking system was used with a tracked probe to collect fiducial points, based on landmarks visible on the skull and in the CT image. This yielded a physical-to-image transform with fiducial registration error (FRE) of 2.3 mm. A brass replica of the hydrophone was then mounted to a custom adapter with a rigid-body tracker. A standard pivot calibration was performed to measure the tip offset (RMSE=0.4 mm). Finally, the tracked hydrophone was mounted to the 3-axis stage and translated over a range of grid points, with the optical tracker recording the tip position at each point. A second rigid registration was performed yielding a transform between motor coordinates and optically tracked coordinates (FRE=0.3 mm). Combining this with the tip offset and the physical-to-image transform yielded a chain of transforms from the motor coordinates of the hydrophone tip into the CT image. This was then used to plan a 3D beam mapping trajectory in a cortical area of the skull piece. Combining the above errors yields a fiducial registration error of 2.4 mm. Though this is an underestimate of the true target registration error, it gives some sense of the error scale involved (West & Maurer 2004). An additional safety margin from the skull interior was therefore applied to trajectory planning. The resulting trajectory had an apparent closest approach of about 3.5mm to account for registration and tracking uncertainty.

The optical tracking system was used in a final step to collect surface points on the interior spherical face of the transducer. These points were used in a least-squares fit to localize the transducer's geometric center within optically-tracked coordinates, and using the transform chain, project transducer elements into the coordinate system of the 3-axis stage. This Cartesian coordinate system was then used to define a 3D simulation grid. Elements were placed in the simulation grid along with the registered CT scan, and simulated using k-Wave (Treeby et al. 2012). Sound speed, density, and attenuation were linearly mapped from CT Hounsfield units, similar to the procedure in Marquet *et al.* (Marquet et al. 2009). The registration did not provide information about alignment and the azimuthal rotation of the transducer about its axis, and simulation setup assumed the sonication axis was parallel to the x-axis and used an arbitrary azimuthal rotation.

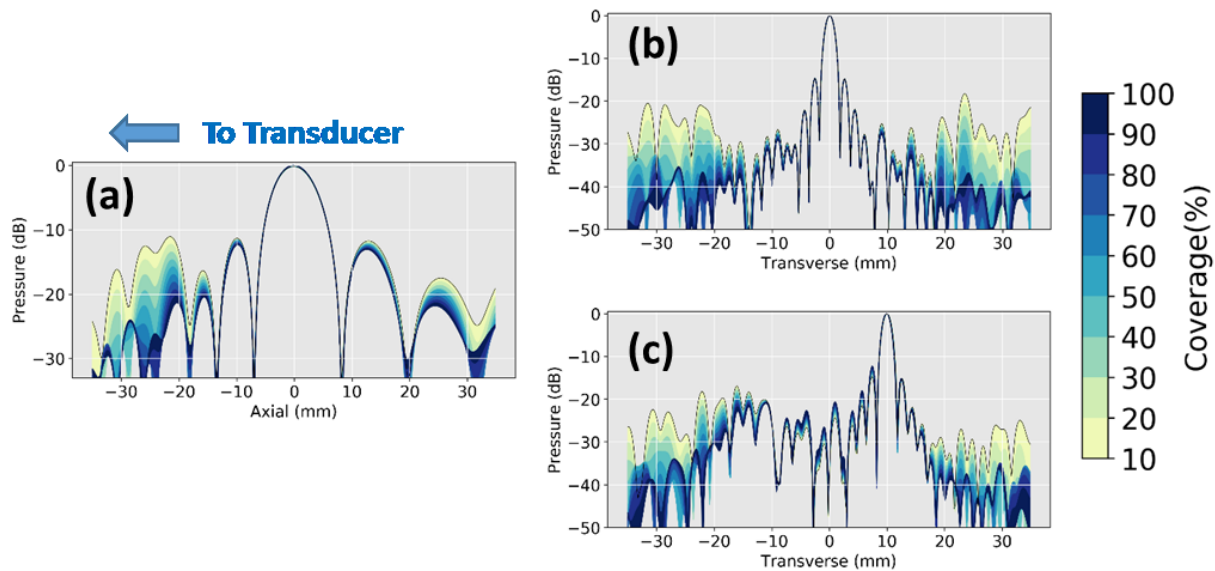

**Figure 2** – Simulated effect of array coverage on grating lobe levels. Coverage % indicates the fraction of cap area that is occupied by active transmit area. **(a)** Beam profile along the sonication axis. **(b)** Transverse profile in the focal plane. **(c)** Same profile as (b), but when the focus is steered 1cm off-axis. These simulations used the final design geometry, ROC=72mm, diameter=103mm, 128 elements, but element sizes were varied and allowed to overlap. The fabricated design has a coverage fraction of 46%.

## Results

*Array coverage percentage and steering*

Simulations of array coverage at the final spherical geometry demonstrate, for the non-steered case, a uniform downward trend in grating lobe and side lobe levels as coverage/density increases from 10% to 100% (**Figure 2**). Color contours correspond to array density. Secondary lobe height decreased with higher array coverage along the axial and transverse profiles (Figure 2a and 2b). Conversely, inter-lobe minima trend to higher pressures with increasing array coverage, demonstrating that larger elements lead to more diffuse acoustic output in peripheral lobe regions. Relative variability due to changing array density decreases towards the central maximum. When the beam is steered off-axis (**Figure 2c**), some maxima see the trend reverse, becoming higher with increasing array density. These results are consistent with previously cited studies about the effects of array density and directivity on grating lobes and steering.

The influence of spherical geometry on focus size can be seen in **Figure 3**, which shows the full-width at 50% and 90% of the pressure maximum (FW50 and FW90, respectively) in the axial and transverse directions. It can be seen from the diagonally oriented contour lines in these figures that--at a fixed frequency and element size--focal size is correlated with f-number (focal length divided by diameter of aperture). Array coverage contours are overlaid, based on the maximum number of elements that fit onto the cap within the stated constraints. We chose a design that produced a small focus, has a convenient focal length with respect to macaque head size, and had nearly the maximum coverage achievable. The chosen design is denoted with a red star and achieves a balance of focus size and area coverage.

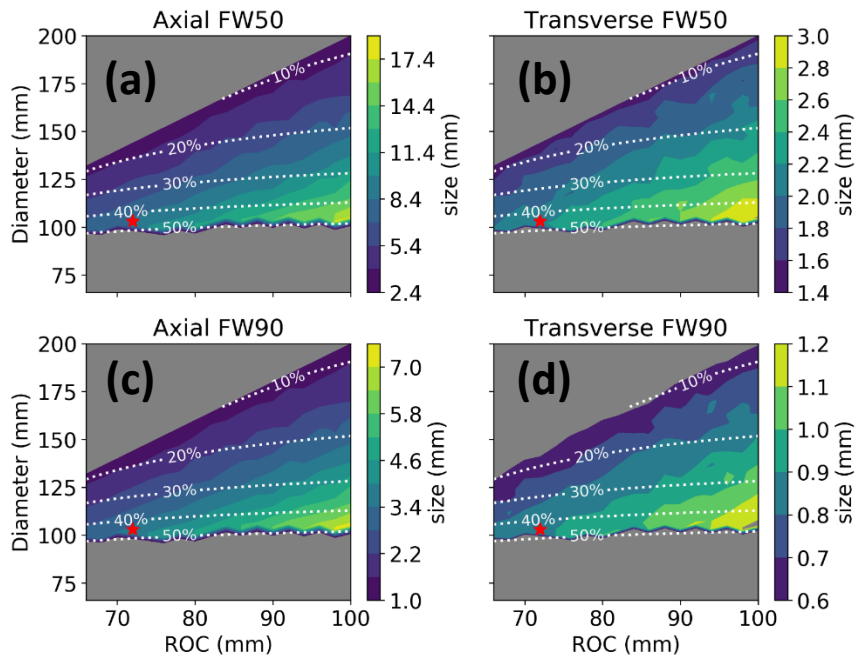

**Figure 3** – Size of the natural focus as a function of spherical geometry, using Rayleigh-Sommerfeld pressure. **(a)-(b):** Full-width half max (FW50). **(c)-(d):** Full-width at 90% max (FW90). **Left column**: size along an axial profile. **Right column**: size along profile in focal plane. Color contours indicate size in each direction. They are diagonally oriented along lines of constant focal ratio (ROC / Diameter), with fluctuations due to randomization of array elements. White contours show the array coverage percent, which is determined by fitting the largest number of elements onto the cap with the stated non-overlap and PCD-avoidance constrains. A red star indicates the chosen design. ROC = Radius of curvature.

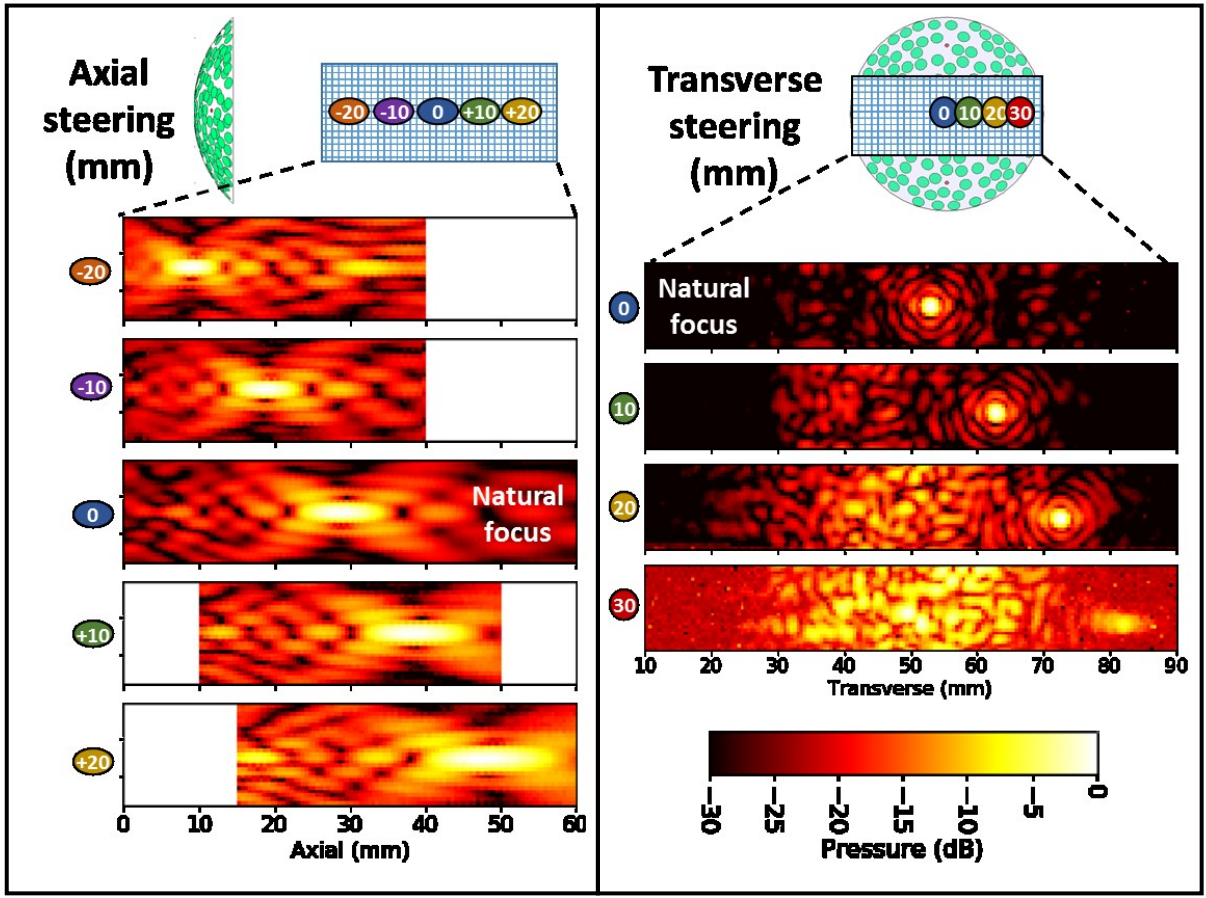

**Figure 4 –** Hydrophone beam maps and grating lobes as the focus is electronically steered. **Left:** steering in 10mm increments along the sonication axis, from -20 to +20 mm. **Right**: steering in the focal plane up to +30 mm. Focus degradation occurs rapidly when steering off-axis.

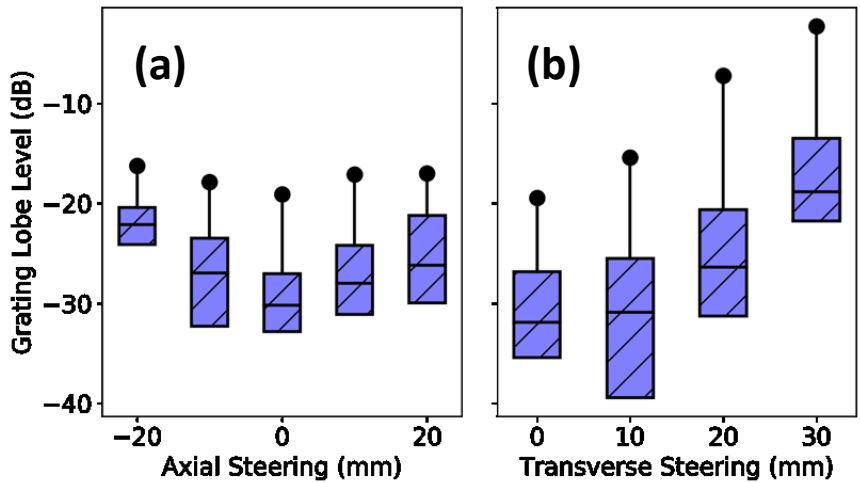

**Figure 5 –** Distributions of grating lobe levels from the data in Figure 4. The boxes represent $25^{th}$, $50^{th}$, $75^{th}$ percentiles, and the max. Pressures are relative to the steered focus in each case. Distributions exclude the region immediately around the focus and first lobe. **(a)** Axial steering **(b)** Transverse steering.

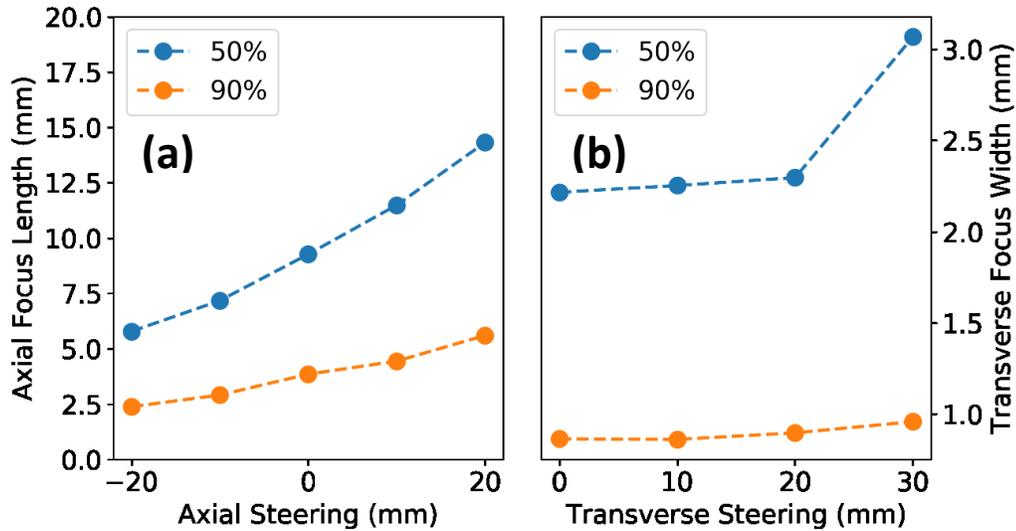

**Figure 6** – Dimensions of the pressure focus measured with needle hydrophone data from Figure 4 as the focus is electronically steered. **(a)** Axial steering **(b)** Transverse steering. '50%' is the full-width half max in each direction. '90%' is the full-width at 0.9 of the max.

*Output measurements*

The chosen design is spherically focused with radius of curvature (ROC) 72 mm and 103 mm diameter at the opening face (f-number 0.7), and a coverage fraction of 46%. The focus size has full-width at half maximum (same as FW50) of 9.3mm and 2.2mm in the axial and lateral directions, respectively. Linear simulations were in good agreement with the low-power hydrophone field measurements. Power calibration resulted in linear pressure output in neuromodulation pressure range, measured waveforms <1MPa at the focus were narrow-band after a short ramp-up (~6 cycles). Pressure output at the maximum input amplitude tested was 12.5 MPa / 8 MPa peak positive / negative. Focal pressure, width, and grating lobe levels in the presence of steering were measured in 10mm steering increments (**Figure 4**). With no steering, the axial grating lobes peaked at –19dB, while the first side lobe surrounding the focus peaked at –9.7 dB. Beam maps collected in the axial plane show no significant pre- or post-focal maxima up to 20mm steering, while steering in the focal plane show focus degradation and grating lobes appearing off-axis. **Figure 5** plots the distribution of maxima, excluding the focus and first sidelobe. The largest lobe level with axial steering was -15dB at -20mm. Transverse steering beyond 10mm exceeded this level. Based on these data we estimate that neuromodulation pulses can be steered within an ellipse +/- 20mm around the focus in the axial direction, and +/- 10mm in the transverse direction, in which the peak free-field grating lobe is -15 dB. **Figure 6** shows focal dimensions in axial and transverse directions, as the beam is steered around the focus.

*Transcranial Evaluation*

**Figure 7** depicts the setup for transcranial beam mapping, image registration, motor trajectory planning. A map of pressure was collected for the transcranial case, and the corresponding free-field maps were collected by simply removing the skull piece. An axial slice through the focus is shown in **Figure 8**. The spatial distribution of peak pressure in the non-homogeneous k-Wave simulations was qualitatively similar. Transcranial beam measurements behind a sample skull showed focal broadening to FW50 of 11.5 and 2.7 mm compared to 9.3 mm and 2.2mm in free field (axial and lateral), a 1.5mm shift of the focus in the axial direction towards the transducer, and peak amplitude 28% of the free-field case **(Figure 9)**. Both the focal broadening and proximal shift are expected. Acoustic transparency of this skull piece at 650kHz agreed with expectation, though it is not necessarily representative of the biological variability encountered with live subjects.

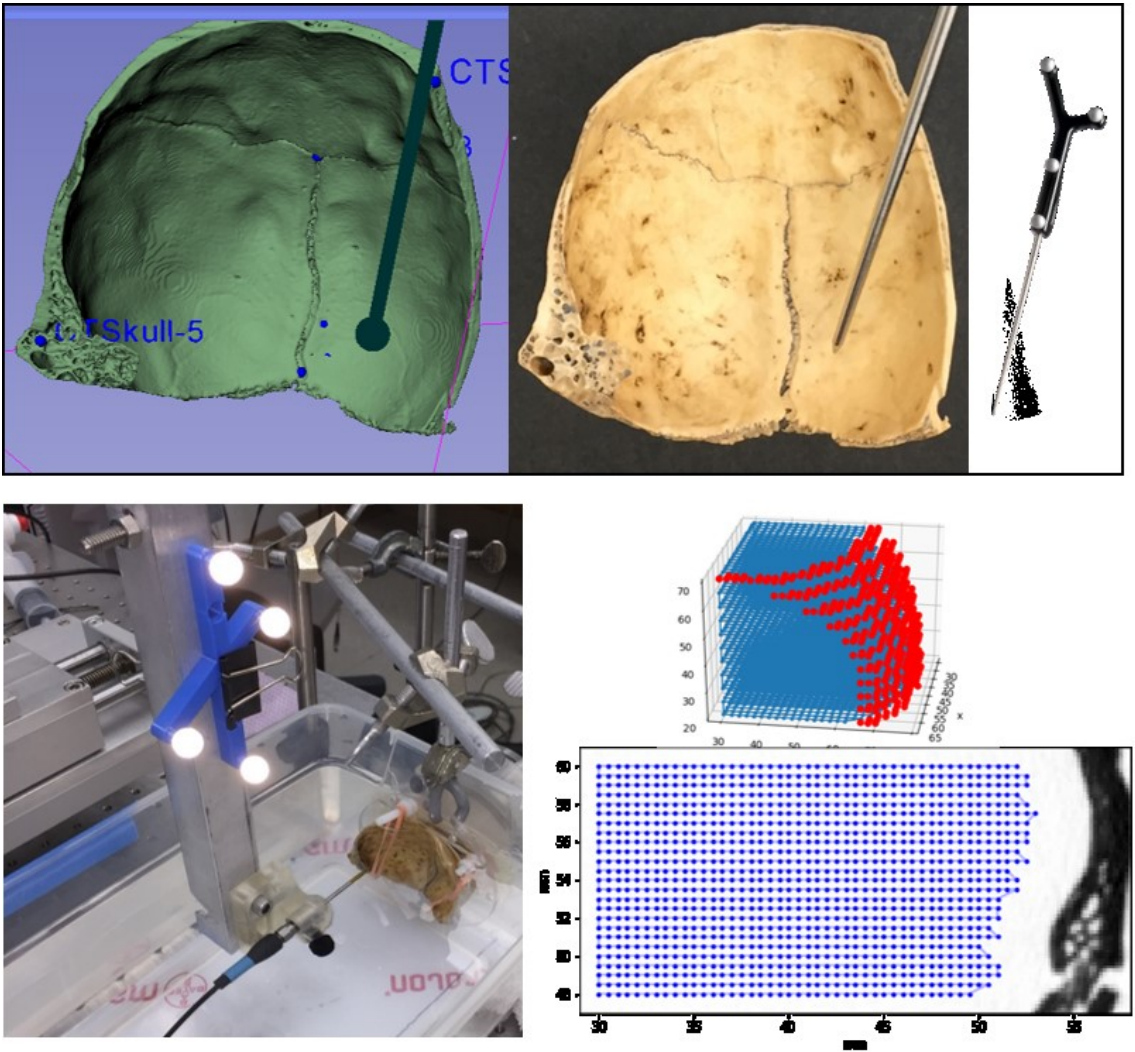

**Figure 7** – Steps involved in guiding the hydrophone tip along the interior of the skull piece using a three-axis stage. Steps include CT-to-physical registration, finding the hydrophone tip offset, physical-to-motor registration, and finally path planning with the projected CT scan. An example hydrophone trajectory is shown on the lower right.

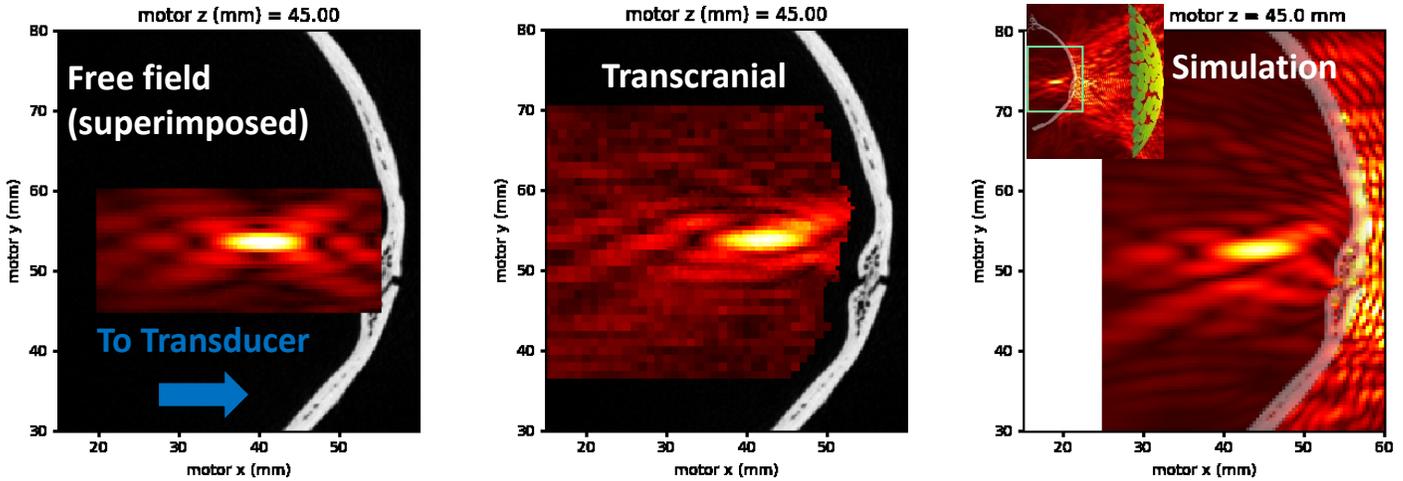

**Figure 8** – Pressure maps in a focal plane slice of the 3D hydrophone collection using the optical tracking setup, showing the free field scan with no skull piece **(left)** and the corresponding scan behind an ex-vivo skull **(center)**. The skull CT was co-registered into motor coordinates and overlaid by interpolating image voxels into the transformed motor frame. For transcranial mapping, a large volume motor scan was collected first with 1mm voxel sides, followed by a smaller high-resolution region around the focus with 0.5mm voxels. The location of the transducer along the sonication axis (approximated to be the motor -x direction) was recorded and used for the K-wave simulation shown **(right)**. Actual azimuthal rotation of the transducer, off-axis alignment, and skull parameters were not known for the simulation.

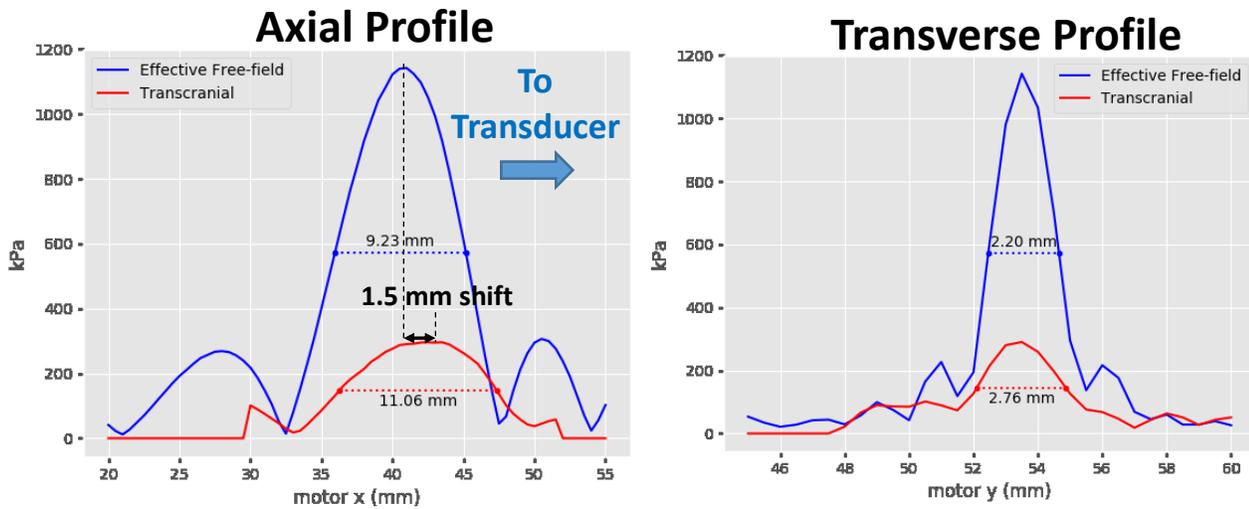

**Figure 9** – Comparison between free-field focus and transcranial focus along the sonication axis **(left)** and in the focal plane **(right)** from hydrophone beam maps. The focus is shifted in the direction of the skull and transducer due to the increased sound speed in the skull, and broadened from phase abberation. Transverse profiles shown on the right coincide with axial focal maxima in both cases. The transcranial pressure was 28% of free-field due to reflection and attenuation by the skull piece.

# Discussion

*Transducer Design*

Using simulations and knowledge gained from prior work, we designed a new randomized sparse spherical array to stimulate the S1 cortex in macaque monkeys. The size and steering constraints for our application relate directly to the geometry and distribution of elements on the transducer array. Our main design objectives were to have a small focus size, low grating lobe levels, ROC appropriate for macaques, and focal steering capability. These design factors were briefly summarized, based on previous work from both volumetric imaging and therapy. In general, focus size and grating lobes are much smaller for arrays with spherical vs. flat construction geometry, but focus degradation with off-axis steering is worse with spherical arrays. Element size and number also influence focus size and steering--a larger number of smaller elements results in focus broadening towards the transducer (Stephens et al. 2011). However, having more elements gives a higher degree of freedom and thus better performance for electronic steering and multi-focusing. For given transducer geometry, grating lobe levels are further modified by the arrangement or spatial distribution of elements. Geometry, element size, count, and arrangement are design variables that must be chosen based on application needs.

*Transcranial Performance*

A sample ex vivo skull section was used to examine skull attenuation and aberration effects on the beam at 650kHz, using a novel approach combining optical tracking, image registration, and beam mapping. The results agree with expectation that the focus is slightly aberrated and shifted toward the transducer. However there are numerous additional details that were not examined which are important with *in vivo* sonication. For example, the specific skull geometry and obliquity of the acoustic axis can aberrate the beam and produce warped focal regions (Hughes et al. 2016). Heating and attenuation through multiple tissue layers are also significant concerns with transcranial FUS, both because of accumulated thermal dose and beam distortion that is detrimental to targeting (Mueller et al. 2016). Attenuation due to muscle layers that surround the skull in macaques was not measured, but is not expected to be a significant risk due to the low pressure involved in neuromodulation. In our experience *in vivo* experiments using a single-element transducer and much smaller focal ratio have shown little risk of skin burn at pressures used for neuromodulation.

Sonication frequency was not considered as a variable, since prior work had identified 650kHz as a safe and effective frequency for transcranial sonication for transcranial thermal lesioning, where higher frequencies are generally preferred because of increased energy absorption (Clement et al. 2000). While it has been shown that 650kHz can stimulate neurons, the exact mechanism has yet to be discovered, and a lower frequency may prove more effective (Kim et al. 2014; Constans et al. 2017). Related, it is not known precisely how the spatial distribution of pressure maps onto the location of modulated neurons. For example, the FWHM contour may not be a meaningful indicator of the modulation threshold. In a study of neuromodulation in a rat where FDG uptake was measured as an indicator for brain activity, the location of FDG-PET signal increase coincided with the estimated 90% contour of the FUS spatial peak, suggesting a threshold-based phenomenon that could be potentially be tuned with specific acoustic intensity profiles (Kim et al. 2013). Functional and *in vitro* studies in the future will be required to resolve this question.

## Conclusions

In this study, we designed and characterized a new randomized, spherically-focused array transducer for transcranial FUS in non-human primates. The transducer and generator system are MR compatible and capable of a large range of pressure output. The focus can be steered over the range of the non-human primate somatosensory cortex and a focal beam can be maintained transcranially. This MR-compatible system will be employed for combined FUS neuromodulation and functional imaging studies.

## Acknowledgements

The authors are grateful for funding support from NIH 5T32EB014841-03, NIH 5R24MH109105, and a Cal-BRAIN Pilot Award.